# How can design help enhance trust calibration in public autonomous vehicles?

Klebanov Yuri, Mikulinsky Romi, Reznikov Tom, Pennington Miles, Suda Yoshihiro, Hiraoka Toshihiro, *Kanzaki Shoichi, IEEE*

*Abstract* — Trust is a multilayered concept with critical relevance when it comes to introducing new technologies. Understanding how humans will interact with complex vehicle systems and preparing for the functional, societal and psychological aspects of autonomous vehicles' entry into our cities is a pressing concern. Design tools can help calibrate the adequate and affordable level of trust needed for a safe and positive experience. This study focuses on passenger interactions capable of enhancing the system trustworthiness and data accuracy in future shared public transportation.

## I. INTRODUCTION

In the past year, an international, multicultural team of designers and technologists were engaged in multidisciplinary collaboration. The team arrived from two prestigious academies:

- A team from the University of Tokyo's Institute of Industrial Science which included members from the University of Tokyo DLX Design Lab and the Advanced Mobility Research Center (ITS Center).

- A team from the Bezalel Academy of Art and Design, Jerusalem.

The research explored trust in autonomous public transportation. Building on the expertise acquired in the Mobility Engineering Labs at the University of Tokyo, the team set out to propose new design-led methods for trust estimation and calibration in autonomous vehicles (AVs). The preliminary motivation was to assess the driver's role in public transportation and to inquire into potential implications of the driver's absence from the vehicle and replacement by a complex robotic system.

## II. RIPPLE - VEHICLE-FLOOR BASED HMI

### A. Methodologies

While in Tokyo, the team started by identifying and developing interactions capable of:

1. Replacing eye contact by means of various human-machine interface (HMI) approaches

2. Learning the passengers' body language, assessing their intentions and responding accordingly in order to meet passenger needs

3. Identifying uncanny interactions and redesigning them into more pleasant experiences capable of reinforcing trust in autonomous systems and agents

4. Hypothesizing that interactions between passengers can be employed to build trust and to counter estrangement and fear of the unknown

During this research, the team utilized design-led methodologies including ideation, design thinking, "sacrificial ideas" and "treasure hunting" (Maudet 2020). The team also commenced a workshop during which they explored a wide range of ideas and suggested and prototyped several collaborative projects and research proposals. One of the selected projects, currently under further development, is Project Ripple – an interactive demo and experimentation tool, currently a prototype of an experimental HMI. In the project's future development, we plan to implement this system on the Navia autonomous shuttle. The future steps of the experiment will take place in both Tokyo and Jerusalem, taking into consideration cultural differences and the responsivity and effectiveness of the HMI.

### B. Principles

The hypothesis is that utilizing the vehicle floor as an interface can help calibrate trust and streamline the implementation and rolling out of AVs. Given that trust is an important factor in the acceptance of autonomous systems that influences using behavior according to Garcia [3] and Hix [5], a supportive user interface is essential – especially in the transition phase towards automated driving, where the "driver" needs to give up control in favor of an unknown feature as Häuslschmid [4] shows. This project does not focus on the driver, but on the passengers' response to autonomous public vehicles, and therein lies its significance and innovation.

Project Ripple's starting point is a single passenger who boards an AV and needs the vehicle to communicate effectively. In the absence of a driver or other passengers, the passenger has no one to ask about the vehicle's route and verify that this is the right vehicle. Ideation, observation and mapping sessions were key steps in the process. Consisting of seven product designers and user interactions designers, the team spent a week at the University of Tokyo ITS. Two days were dedicated to "treasure hunting", as explained in Maudet [8], in mobility and other technological research labs. Two more days were dedicated to ideation sessions, building on the insights of the "treasure hunt".

The ideation sessions yielded several dozen of ideas, scenarios and research proposals. Two leading proposals were selected, one of them being Project Ripple. To better understand the design challenges of the project, another day was dedicated to observing passenger behavior in public transport around the Kashiwa city area, on both autonomous vehicles (U.Tokyo shuttle bus) and human-operated buses, taxis and trains. Drawing on the ideation sessions and the observations, the team concluded that the following aspects would be impacted with the transition to AVs:

- A sense of familiarity, trust and control

- A sense of authority and responsibility inside the vehicle

[1]*Research supported by the JSPS KAKENHI Grants.

Y. Klebanov is with the University of Tokyo Institute of Industrial Science, Tokyo, Japan (+81-3-5452-6164; yurikleb@iis.u-tokyo.ac.jp).

R. Mikulinsky is with Bezalel the Academy of Art and Design, Jerusalem, Israel. (+972-2-6332996; rominska@post.bezalel.ac.il).

- A sense of safety and coherence

The needs identified above – central to this research – derive from the driver's absence: the passenger has no one to confirm safe boarding and alighting; to ask for directions, inquire about payment, or delays and their reasons, to provide guidance during emergencies and so on. The project proposes that by providing such information via new interactive platforms, the vehicle can instill confidence and orientation, and thereby build passenger trust.

The team then started defining the research questions (referred to as "design challenges" in design methodologies) and moved on to validating them through prototyping (see Hix [5] and Sirkin and Ju [11] for additional information about interface design and usability testing). Three scenarios were selected for testing an interaction system that meets the above needs and promotes the following values:

- The vehicle acknowledges the passenger' presence and provides them with information. This interaction is also meant to build and calibrate trust.
- The vehicle itself plays a part in the communication between the passengers and the driverless vehicle. It becomes a mediator and a platform for communicating with the passengers and holds the potential of obviating the need for other interaction platforms (such as smartphones).

*C. Scenarios*

In design, it is commonplace to define specific user scenarios to commence an ideation process that meets specific needs. To address the needs recognized in the previous session, the team suggested providing constant feedback and information to maintain communication with the passenger throughout the entire journey. The team divided the passenger journey into three scenarios. Each focused on a different phase of the journey, allowing the team to evaluate the amount of trust lost or gained and provide more accurate feedback via various interfaces and interactions, creating a more coherent journey experience.

- **Scenario 1 - Boarding and Alighting:** The passenger receives acknowledgement of their presence by a physical and visual gesture upon embarking – the floor acts as a platform to greet the passenger and replace the traditional nod by the driver. Additional gestures notify the passenger as the bus starts moving, breaks or makes any other unexpected movement (Figure 1), instilling a sense of trust, familiarity and safety.
- **Scenario 2 - Detailed Ride Information:** The passenger receives personalized detailed information about the ride and the status of the bus such as stops, final destination, fares, traffic conditions, points of interest, etc. (Figure 2), instilling a sense of trust and enabling the passenger to feel in greater control of the situation.
- **Scenario 3 - Ambient Environmental Information:** The passenger is presented with ambient/background information. This enables them to concentrate on other things such as their phone or book. This would replace the driver in providing safety information such as "watch your step", "don't slip", "please move down the aisle". This would instill a sense of authority, responsibility and safety.

*D. Interfaces*

Several interface design opportunities were explored. The team researched a range of currently popular interfaces such as screens and LED displays installed inside and outside autonomous vehicles, as well as other solutions such as virtual and robotic agents. The team explored and identified multiple surfaces in public vehicles as new opportunities that are not yet in broad use to implement interactive interfaces. For private information, the handlebars and hand rings were proposed. The floor was selected for public information and interactions (Figure 3).

As an experimental approach, the team suggested using natural inspiration for some parts of the interfaces to diverge from today's popular informative anthropomorphic trend design solutions (see: Ruijten [10] and Waytz [12]), that employ a visual language which can be identified as "hi-tech" and therefore may augment estrangement (Figure 4). After prototyping several proposals, the team decided to focus on the interactive floor interface for the first phase of project development.

III. FUTURE STEPS

The above scenarios are currently tested in the lab, using a "prototype booth" which mimics the interior of a bus using interactive flooring, a light detection and ranging (LIDAR) sensor as well as pressure and contact sensors (Figure 5). A small group of participants will be presented with the above scenarios and interact with the prototype. Their interactions will be observed and they will be asked to evaluate and rate them via a questionnaire and interviews.

Based on the insights from the "booth" experiment, the interface and scenario implementation will be refined and deployed in an autonomous Navia shuttle. The following steps will be implemented "in the wild" to be tested with a wider range of users, to test performance in a real-world scenario. Here too questionnaires and interviews will be used, alongside video recordings of user behavior.

The experiment is designed to determine whether the timing, nature and adequacy of information contribute to building trust or whether redundant, inadequate or irrelevant information may affect trust negatively. Given that trust is a dynamic process - as Choi and Ji [2], Lee and See [6], Hoff and Bashir [7] and Noah [9] show - uncertainty may harm trust, and therefore one of the experiment's goals is to learn whether trust can be regained after being lost.

The experiment will be replicated in Israel after the trial in Tokyo and the results of the two experiments will be compared and assessed.

APPENDIX

*Figure 1.* Onboarding acknowledgement feedback

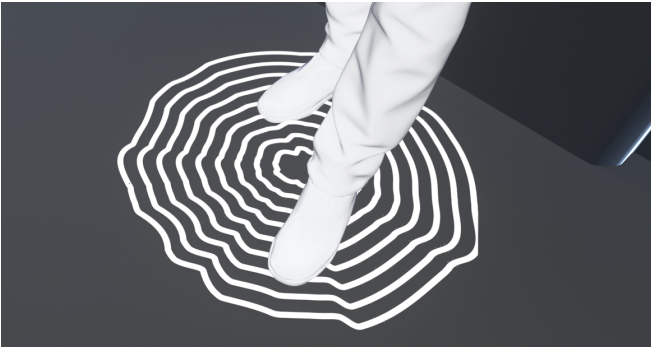

*Figure 2.* Detailed Ride Information

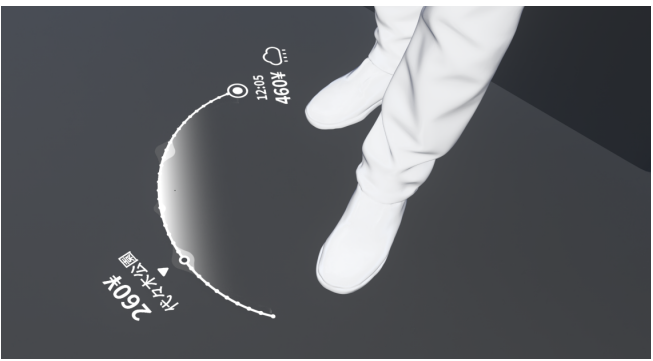

*Figure 3.* Exploration of Interface opportunities in public transport

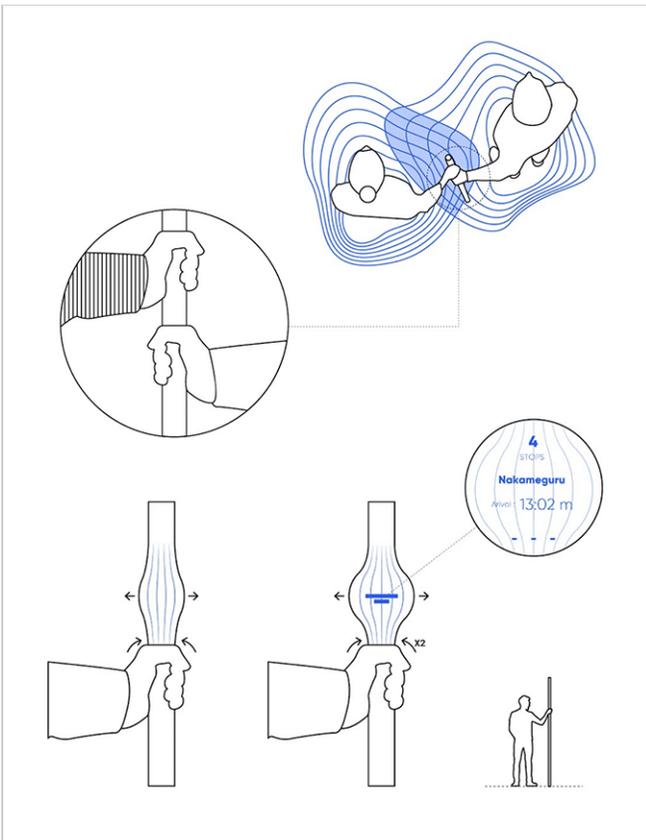

*Figure 4.* Nature-inspired interfaces

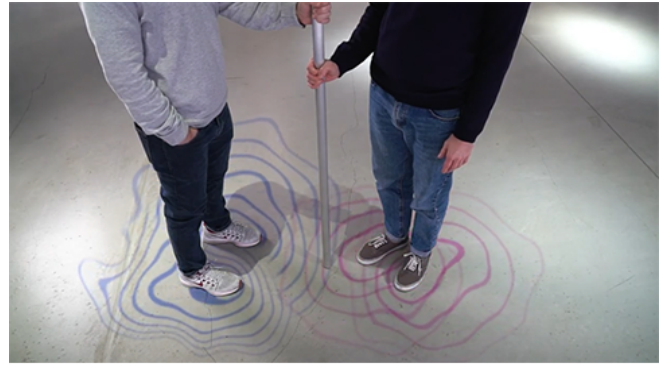

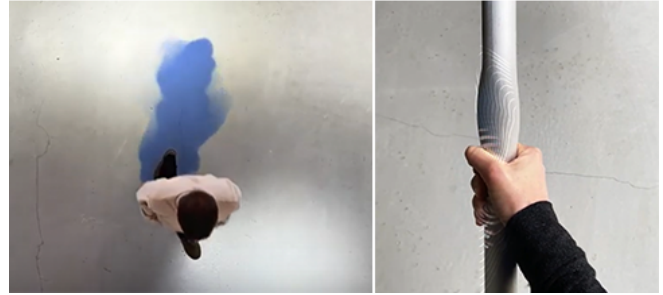

*Figure 5.* Prototyping Booth

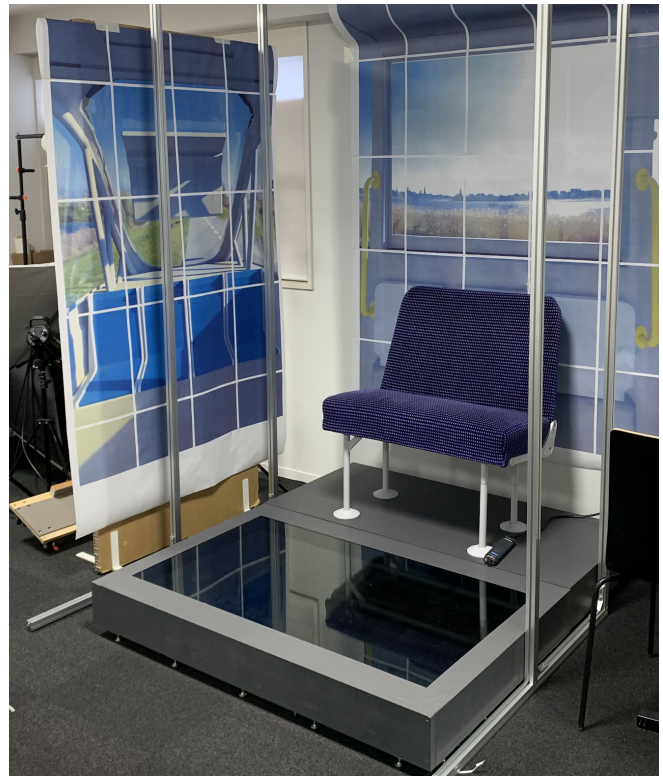

ACKNOWLEDGMENT

The authors would like to thank all the participants in Israel and Japan. The workshop took place thanks to the kind support of Bezalel's Research and Innovation Authority.REFERENCES

[1] O. Carsten, and M. H Martens, "How can humans understand their automated cars? HMI principles, problems and solutions. *Cognition, Technology & Work,* 21(1), 2019, pp. 3-20.

[2] J. K. Choi, and Y. G. Ji, "Investigating the importance of trust on adopting an autonomous vehicle," I*nternational Journal of Human–Computer Interaction,* Volume 31, (10) 2015, pp. 692-702

[3] D. Garcia, C. Kreutzer, K. Badillo-Urquiola and M. Mouloua "Measuring trust of Autonomous Vehicles: A development and validation study," *Proceedings of HCI International 2015, Part II* -Posters' Extended Abstracts (HCI International 2015), Constantine Stephanidis (Ed.). Springer International Publishing, Cham, 2015, pp. 610–615.

[4] R. Häuslschmid, M. Von Buelow, B. Pfleging, and A. Buts. "Supporting trust in autonomous driving," *Proceedings of International Conference on Intelligent User Interfaces, IUI 2017,* March 13 - 16, 2017, pp. 319-329. doi: 10.1145/3025171.3025198

[5] D. Hix and H. Hartson*, Developing user interfaces: ensuring usability through product and process*, Wiley, New York. 1993

[6] K. A. Hoff, and M. Bashir, "Trust in automation: integrating empirical evidence on factors that influence trust." *Hum Factors*, 57(3), 2015, pp. 407-434.

[7] J. D. Lee and K. A. See, "Trust in automation: designing for appropriate reliance." Human factors, 46(1), 2004, pp. 50-80.

[8] N. Maudet, S. Asada, and M. Pennington, M, "Treasure Hunting: an exploratory study of how designers and scientists identify potential collaborative projects," in S. Boess, M. Cheung, and R. Cain, (Eds.), Synergy - DRS International Conference 2020, 11-14 August, 2020.

[9] B. E. Noah, P. Wintersberger, A. G. Mirnig, S. Thakkar, F. Yan, T. M. Gable, and R. McCall, "First workshop on trust in the age of automated driving" in *Proceedings of the 9th International Conference on Automotive User Interfaces and Interactive Vehicular Applications Adjunct - Automotive UI '17,* Oldenburg, Germany. September 24-27, 2017. doi: 10.1145/3131726.3131729

[10] P.A.M Ruijten, J.M.B. Terken, and S.N Chandramouli, "Enhancing trust in Autonomous Vehicles through intelligent user interfaces that mimic human behavior," M*ultimodal Technol.* Interaction 2, no. 4: 62. 2018, doi: 10.3390/mti2040062.

[11] D. Sirkin and W. Ju, "Embodied Design Improvisation: A Method to Make Tacit Design Knowledge Explicit and Usable," in *Design Thinking Research. Understanding Innovation*. H. Plattner, C. Meinel, L. Leifer, Eds. Springer, Cham. 2015, pp. 195-209 doi: 10.1007/978-3-319-06823-7_11

[12] A. Waytz, J. Heafner., N. Epley, "The mind in the machine: Anthropomorphism increases trust in an autonomous vehicle, *Journal of Experimental Social Psychology*, Volume 52, 2014, pp. 113-117. doi: 10.1016/j.jesp.2014.01.005